\documentclass[prd,amsmath,amssymb,showpacs,superscriptaddress,nofootinbib,twocolumn]{revtex4-1}
\usepackage{graphicx}
\usepackage{epsfig,graphics,subfigure,psfrag,amsmath,amssymb}
\usepackage{lineno}
\usepackage{dcolumn}
\usepackage{bm}
\usepackage{overpic}
\usepackage{xspace}

\usepackage{color}
\usepackage{multirow}
\usepackage{colortbl}
\usepackage{epstopdf}
\usepackage[colorlinks,linkcolor=blue,anchorcolor=blue,citecolor=blue]{hyperref}

\newcommand{\BR}{{\cal B}}

\newcommand{\jpsi}{J/\psi}

\newcommand{\pip}{\pi^{+}}
\newcommand{\pim}{\pi^{-}}
\newcommand{\etap}{\eta^{\prime}}

\newcommand{\kp}{K^{+}}
\newcommand{\km}{K^{-}}
\newcommand{\kk}{K^{+}K^{-}}

\newcommand{\BESIII}{BES\uppercase\expandafter{\romannumeral3}\xspace}

\begin{document}

\title{\texorpdfstring{\boldmath Search for the weak decay $\etap\to K^{\pm}\pi^{\mp}$ and precise measurement of the branching fraction $\BR(\jpsi\to\phi\etap)$}{Search for eta' to Kpi}}

\author{M.~Ablikim$^{1}$, M.~N.~Achasov$^{9,f}$, X.~C.~Ai$^{1}$, O.~Albayrak$^{5}$, M.~Albrecht$^{4}$, D.~J.~Ambrose$^{44}$, A.~Amoroso$^{49A,49C}$, F.~F.~An$^{1}$, Q.~An$^{46,a}$, J.~Z.~Bai$^{1}$
% ...
\\
\vspace{0.2cm}
(BESIII Collaboration)\\
\vspace{0.2cm} {\it
$^{1}$ Institute of High Energy Physics, Beijing 100049, People's Republic of China\\
% ...
$^{4}$ Bochum Ruhr-University, D-44780 Bochum, Germany\\
$^{5}$ Carnegie Mellon University, Pittsburgh, Pennsylvania 15213, USA\\
% ...
$^{9}$ G.I. Budker Institute of Nuclear Physics SB RAS (BINP), Novosibirsk 630090, Russia\\
% ...
$^{44}$ University of Rochester, Rochester, New York 14627, USA\\
% ...
$^{46}$ University of Science and Technology of China, Hefei 230026, People's Republic of China\\
% ...
$^{49}$ (A)University of Turin, I-10125, Turin, Italy; (B)University of Eastern Piedmont, I-15121, Alessandria, Italy; (C)INFN, I-10125, Turin, Italy\\
% ...
$^{a}$ Also at State Key Laboratory of Particle Detection and Electronics, Beijing 100049, Hefei 230026, People's Republic of China\\
% ...
$^{f}$ Also at the Novosibirsk State University, Novosibirsk, 630090, Russia\\
% ...
}
}

\date{\today}


\begin{abstract}
  We present the first search for the rare decay of $\eta^\prime$ into
  $K^{\pm}\pi^{\mp}$ in $\jpsi\to\phi\etap$, using a sample of
  $1.3\times10^{9}$ $\jpsi$ events collected with the BESIII
  detector. No significant signal is observed, and the upper limit at
  the 90\% confidence level for the ratio $\frac{\BR(\eta'\to
    K^{\pm}\pi^{\mp})}{\BR(\eta'\to
    \gamma\pi^{+}\pi^{-})}$ is determined to be
  $1.3\times10^{-4}$. In addition, we report the measurement of the branching
  fraction of $\jpsi\to
  \phi\eta'$ to be $(5.10
  \pm0.03(\text{stat.})\pm0.32(\text{syst.})) \times
  10^{-4}$, which agrees with previous results from BESII.
\end{abstract}

\maketitle

\section{Introduction}
Non-leptonic weak decays are valuable tools for testing the Standard
Model (SM), the Kobayashi-Maskawa (KM) mechanism, and the unitarity of
the Cabibbo-Kobayashi-Maskawa (CKM) matrix, and for exploring physics
beyond the SM. Among non-leptonic decays, the decay of the light
pseudoscalar meson $\eta'\to K^{\pm}\pi^{\mp}$ is interesting because
it is fundamental to understand the long-standing problem of the
$\Delta I=1/2$ rule in weak non-leptonic interactions.

The experimental $\Delta I=1/2$ rule was first established in the
decay $K\to\pi\pi$. A neutral kaon may decay into two pions with
amplitudes $A_{0}$ or $A_{2}$, respectively. As the real parts of these
amplitudes, Re$A_{0}$ is dominated by $\Delta I=1/2$ transitions and
Re$A_{2}$ receives contributions from $\Delta I=3/2$ transitions, the
former transitions dominate Re$A_{0}$, which expresses the so-called
$\Delta I=1/2$ rule~\cite{deltaIrule1,deltaIrule2}
\begin{equation}
  \begin{aligned}
    \frac{\text{Re}A_{0}}{\text{Re}A_{2}}=22.35.
  \end{aligned}
\end{equation}
Despite nearly 50 years of efforts, the microscopic dynamical
mechanism responsible for such a striking phenomenon is still elusive.
The decay $\eta'\to K^{\pm}\pi^{\mp}$ receives contributions from both
the $\Delta I=1/2$ and $\Delta I=3/2$ parts of the weak
hamiltonian~\cite{bergstrom1988weak}. It is possible to see whether
the $\Delta I=1/2$ rule is functional in this type of decay, and this
could shed light on the origin of this rule.  The branching fraction
of $\etap\to K^{\pm}\pi^{\mp}$ decay is predicted to be of the order
of $10^{-10}$ or higher~\cite{bergstrom1988weak}, with a large
long-range hadronic contribution expected, which should become
observable in high luminosity electron-positron collisions.

At present, there is no experimental information on the decay
$\etap\to K^\pm\pi^\mp$.  The world's largest sample of
$1.3\times 10^9$ $\jpsi$ events produced at rest and collected with
the \BESIII detector therefore offers a good opportunity to search for this rare
decay. In this paper, the measurement of the ratio
$\frac{\BR(\eta'\to K^{\pm}\pi^{\mp})}{\BR(\eta'\to
  \gamma\pi^{+}\pi^{-})}$ is presented, where the $\eta^\prime$ is
produced in the decay $J/\psi\to\phi\eta^\prime$. The advantage of
comparing these two $\eta^\prime$ decay channels is that parts of the
systematic uncertainties due to the tracking, the particle
identification (PID), the branching fractions $\BR(J/\psi\to\phi\etap)$ and $\BR(\phi\to K^{+}K^{-})$,
and the number of $J/\psi$ events cancel in
the ratio. A measurement of the branching fraction $J/\psi\to\phi\etap$
is also presented in which $\phi$ is reconstructed in its $\kk$
decay mode and $\etap$ is detected in the $\gamma\pi^+\pi^-$ decay
mode. This can be compared with the results reported by the
BESII~\cite{ablikim2005measurements},
MarkIII~\cite{MARKIIImeasurements}, and DM2~\cite{DM2measurements} collaborations.

\section{\texorpdfstring{Detector and Monte Carlo Simulation}{Detector and MC Simulation}}
BEPCII is a double-ring $e^{+}e^{-}$ collider designed to provide a
peak luminosity of $10^{33}$~cm$^{-2}s^{-1}$ at the center-of-mass
(c.m.) energy of 3.770 GeV.  The \BESIII~\cite{Ablikim2010345}
detector, with a geometrical acceptance of 93\% of the 4$\pi$ stereo
angle, is operating in a magnetic field of 1.0 T provided by a
superconducting solenoid magnet. It is composed of a helium-based
drift chamber (MDC), a plastic scintillator Time-Of-Flight (TOF)
system, a CsI(Tl) electromagnetic Calorimeter (EMC) and a multi-layer
resistive plate chamber (RPC) muon counter system (MUC).

Monte Carlo (MC) simulations are used to determine the mass resolutions
and detection efficiencies. The GEANT4-based simulation
software BOOST~\cite{ref:boost} includes the geometric and material
description of the \BESIII detector, the detector response, and the
digitization models, as well as the detector running conditions and
performance.  The production of the $\jpsi$ resonance is simulated with
the MC event generator KKMC~\cite{ref:kkmc,ref:kkmc2}, while the
decays are generated by EVTGEN~\cite{ref:evtgen} for known decay modes
with branching fractions set to the Particle Data Group
(PDG)~\cite{PDG2014} world average values, and by
LUNDCHARM~\cite{ref:lundcharm} for the remaining unknown decays.  The
analysis is performed in the framework of the \BESIII offline software
system (BOSS)~\cite{ref:boss}.

\section{Data analysis}
\subsection{\texorpdfstring{ $\jpsi\to\phi\eta'$, $\eta^\prime\to\gamma\pi^+\pi^-$}{eta' to gampipi}}
For the decay $\jpsi\to\phi\eta'$, $\phi\to\kp\km$,
$\eta^\prime\to\gamma\pi^+\pi^-$, candidate events are selected
by requiring four well reconstructed charged tracks and at least one
isolated photon in the EMC.  The four charged tracks are required to
have zero net charge. Each charged track, reconstructed using hits in
the MDC, is required to be in the polar angle range
$|\cos\theta| < 0.93$ and pass within $\pm10$~cm of the interaction
point along the beam direction, and within $\pm1$~cm in the plane
perpendicular to the beam, with respect to the interaction point. For
each charged track, information from the TOF and the specific
ionization measured in the MDC ($dE/dx$) are combined to
form PID confidence levels (C.L.) for the $K$, $\pi$ and $p$
hypotheses, and the particle type with the highest C.L.\ is assigned to
each track. Two of the tracks are required to be identified as kaons and the
remaining two tracks as pions.

Photon candidates are reconstructed by clusters of energy deposited in
the EMC.  The energy deposited in the TOF counter in front of the EMC
is included to improve the reconstruction efficiency and the energy
resolution. Photon candidates are required to have a deposited
energy larger than 25 MeV in the barrel region ($|\cos\theta|<0.80$)
and 50 MeV in the end-cap region ($0.86<|\cos\theta|<0.92$). EMC
cluster timing requirements are used to suppress electronic noise and
energy deposits that are unrelated to the event. To eliminate showers
associated with charged particles, the angle between the cluster and
the nearest track must be larger than 15$^{\circ}$.
\begin{figure}[htbp]
  \includegraphics[width=0.45\textwidth]{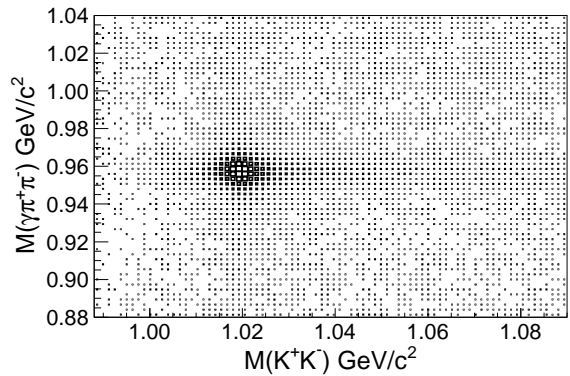}
  \caption{\label{PhiGamPiPi_scatter}Scatter plot of $M(\gamma\pip\pim)$ versus $M(\kk)$ .}
\end{figure}
\begin{figure*}[htbp]
  \includegraphics[width=0.45\textwidth]{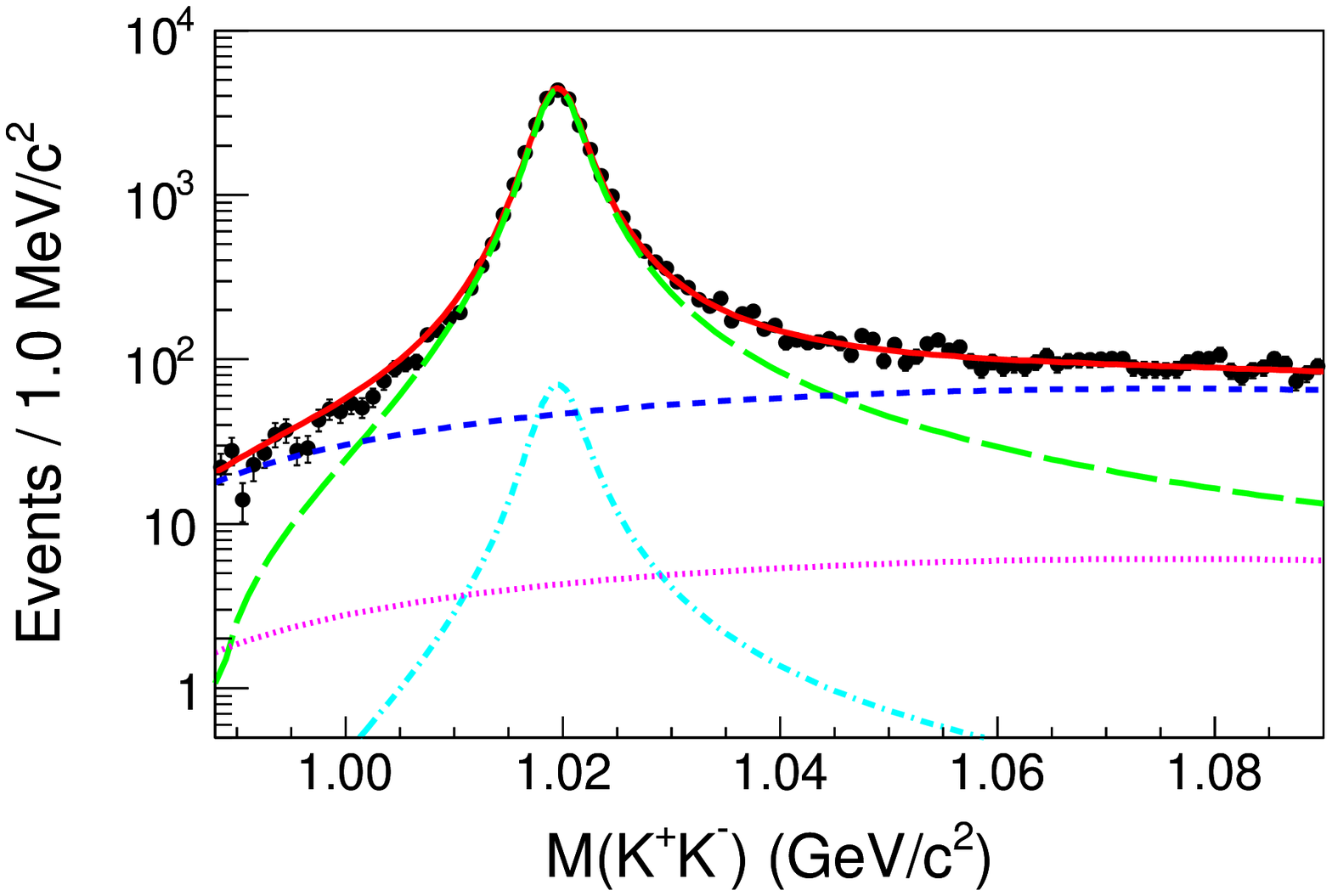}\put(-70,110){\bf (a)}
  \includegraphics[width=0.45\textwidth]{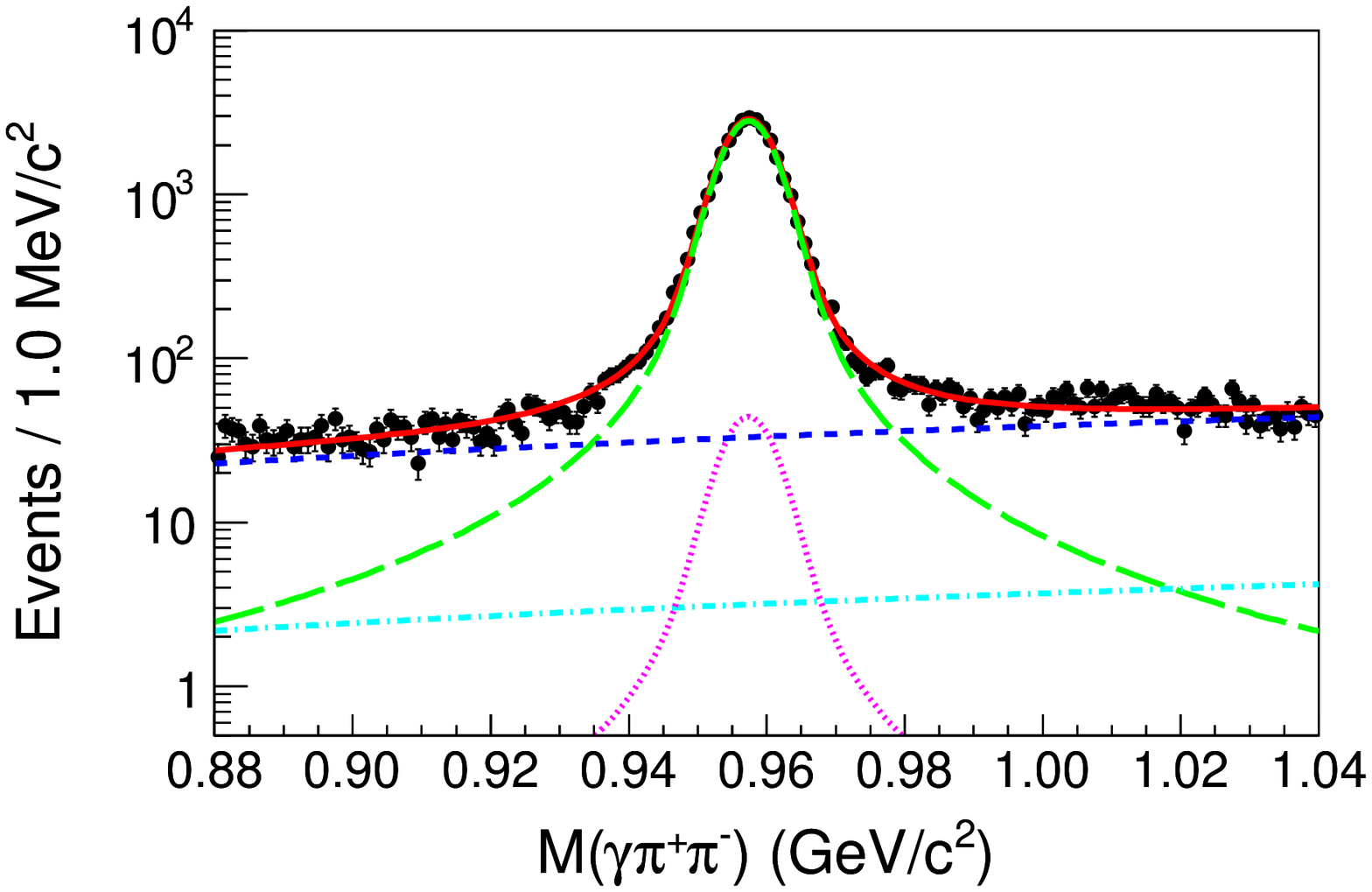}\put(-70,110){\bf (b)}
  \caption{\label{nominal_fit_result}Distributions of (a) $M(\kk)$ and
    (b) $M(\gamma\pip\pim)$ with projections of the fit result superimposed
    for $\jpsi\to\phi\eta', \phi\to\kk, \eta'\to\gamma\pip\pim$. The
    dots with errors are for data, the solid curve shows the result of
    the fit to signal plus background distributions, the long-dashed
    curve is for $\phi\eta'$ signal, the dot-dashed curve shows the
    non-$\eta'$-peaking background, the dotted curve shows the
    non-$\phi$-peaking background, and the short-dashed curve is for
    non-$\phi\eta'$ background.}
\end{figure*}

A four-constraint (4C) kinematic fit is performed to the
$\gamma\kk\pip\pim$ hypothesis. For events with more than one photon
candidate, the candidate combination with the smallest $\chi^{2}_{4C}$
is selected, and it is required that $\chi^{2}_{4C}<50$.

The scatter plot of $M(\gamma\pi^+\pi^-)$ versus $M(\kk)$ is shown in
Fig.~\ref{PhiGamPiPi_scatter}, where the $J/\psi\to\phi\eta^\prime$
decay is clearly visible. To extract the number of $\phi\eta^\prime$ events,
an unbinned extended maximum likelihood fit is performed to the
$M(\gamma\pip\pim)$ versus $M(\kk)$ distribution  with the
requirements of 0.988~GeV/$c^2$ $< M(\kk) < 1.090$~GeV/$c^2$ and
0.880~GeV/$c^2$ $< M(\gamma\pip\pim) < 1.040$~GeV/$c^2$. Assuming zero
correlation between the two discriminating variables $M(\kk)$ and
$M(\gamma\pip\pim)$, the composite probability density function (PDF)
in the 2-dimensional fit is constructed as follows
\begin{equation}
  \begin{aligned}
    F&=N_\text{sig}\times (F_\text{sig}^{\phi}\cdot F_\text{sig}^{\eta'})\\
    &+N_\text{bkg}^{\text{non-}\eta'}\times (F_\text{sig}^{\phi}\cdot F_\text{bkg}^{\text{non-}\eta'})\\
    &+N_\text{bkg}^{\text{non-}\phi}\times (F_\text{bkg}^{\text{non-}\phi}\cdot F_\text{sig}^{\eta'})\\
    &+N_\text{bkg}^{\text{non-}\phi\eta'}\times (F_\text{bkg}^{\text{non-}\phi}\cdot F_\text{bkg}^{\text{non-}\eta'}).
  \end{aligned}
\end{equation}
Here, the signal shape for $\phi$ (\emph{i.e.} $F_\text{sig}^{\phi}$) is modeled
with a relativistic Breit-Wigner function convoluted with a Gaussian
function taking into account the detector resolution; the signal
shape for $\eta'$ ($i.e.~F_\text{sig}^{\eta'}$) is described by a normal
Breit-Wigner function convoluted with a Gaussian function. The widths
and masses of $\phi$ and $\eta'$ are free parameters in the fit. The
background shape of $\phi$ ($F_\text{bkg}^{\text{non-}\phi}$) is described by
a second order Chebychev polynomial function, and the background shape
of $\eta'$ ($F_\text{bkg}^{\text{non-}\eta'}$) is described by a first order
Chebychev polynomial function. All parameters related to the
background shapes are free in the fit. $N_\text{sig}$ is the number of
$\jpsi\to\phi\eta', \phi\to\kk, \eta'\to\gamma\pip\pim$ signal
events. The backgrounds are divided into three categories:
non-$\phi\eta'$ background ($i.e.~\jpsi\to\gamma\kk\pip\pim$);
non-$\phi$-peaking background ($i.e.~\jpsi\to\kk\eta'$); and
non-$\eta'$-peaking background
($i.e.~\jpsi\to\phi\gamma\pip\pim$). The parameters $N_\text{bkg}^{\text{non-}\phi\eta'}$,
$N_\text{bkg}^{\text{non-}\phi}$ and $N_\text{bkg}^{\text{non-}\eta'}$ are the
corresponding three background yields.

The resulting fitted number of signal events is
$N_\text{sig}=(31321\pm201)$; the projections of the fit on the $M(\kk)$
and $M(\gamma\pip\pim)$ distributions are shown in
Figs.~\ref{nominal_fit_result} (a) and (b), respectively.  The
detection efficiency, $32.96\pm0.04$\%, is obtained from the MC
simulation in which the angular distribution and the shape of
$M(\pi^{+}\pi^{-})$ are taken into account according to a
previous \BESIII measurement for
$\eta'\to\pi^{+}\pi^{-}e^{+}e^{-}$~\cite{Eta'ToPiPiEE}, where the
non-resonant contribution (known as the ``box anomaly'') is included
in the simulation of $\etap\to\gamma\pip\pim$.

\subsection{\texorpdfstring{$J/\psi\to\phi\eta', \eta'\to K^{\pm}\pi^{\mp}$}{eta' to Kpi}}
\begin{figure*}[htbp]
  \includegraphics[width=0.45\textwidth]{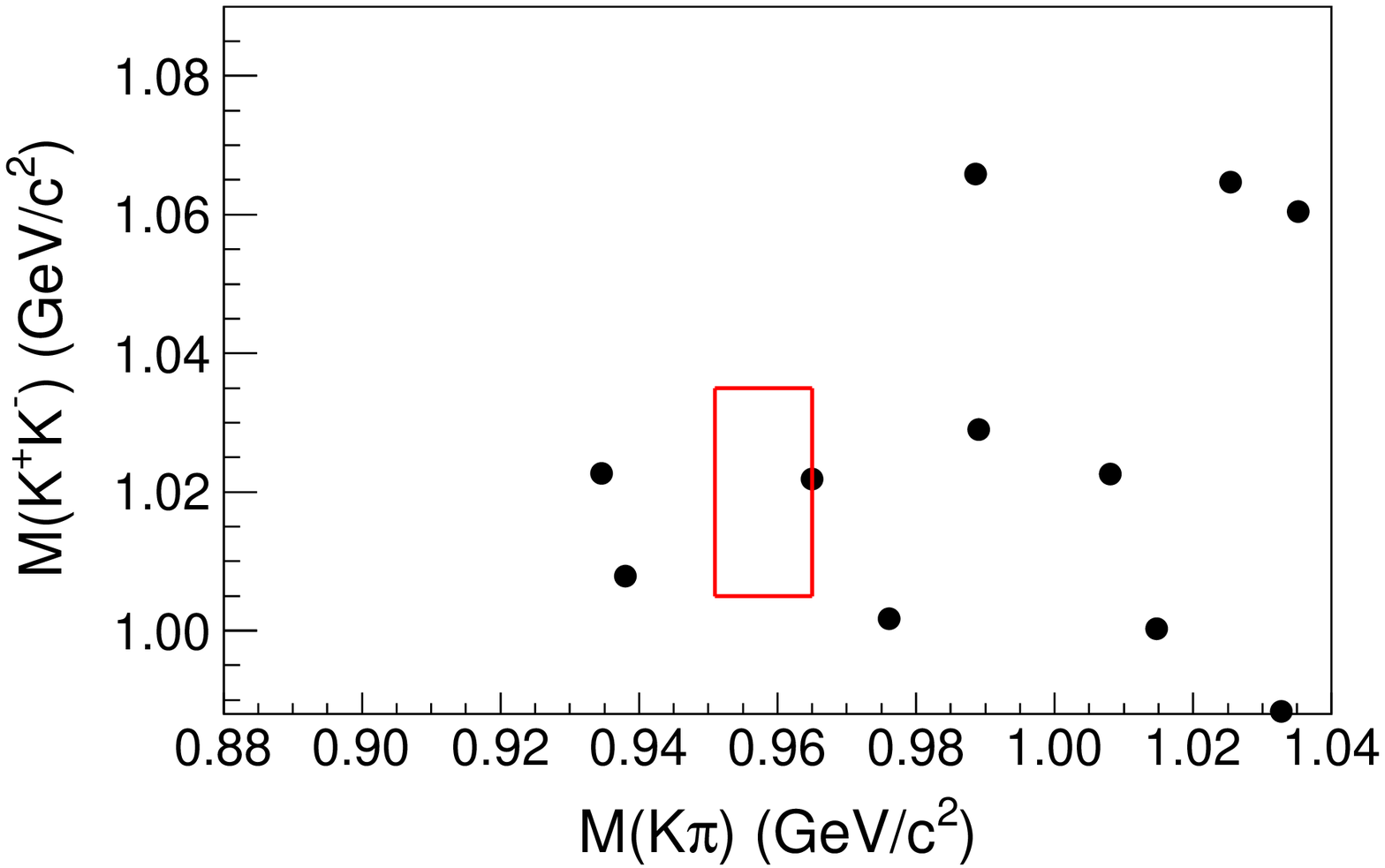}\put(-50,110){\bf (a)}
  \includegraphics[width=0.45\textwidth]{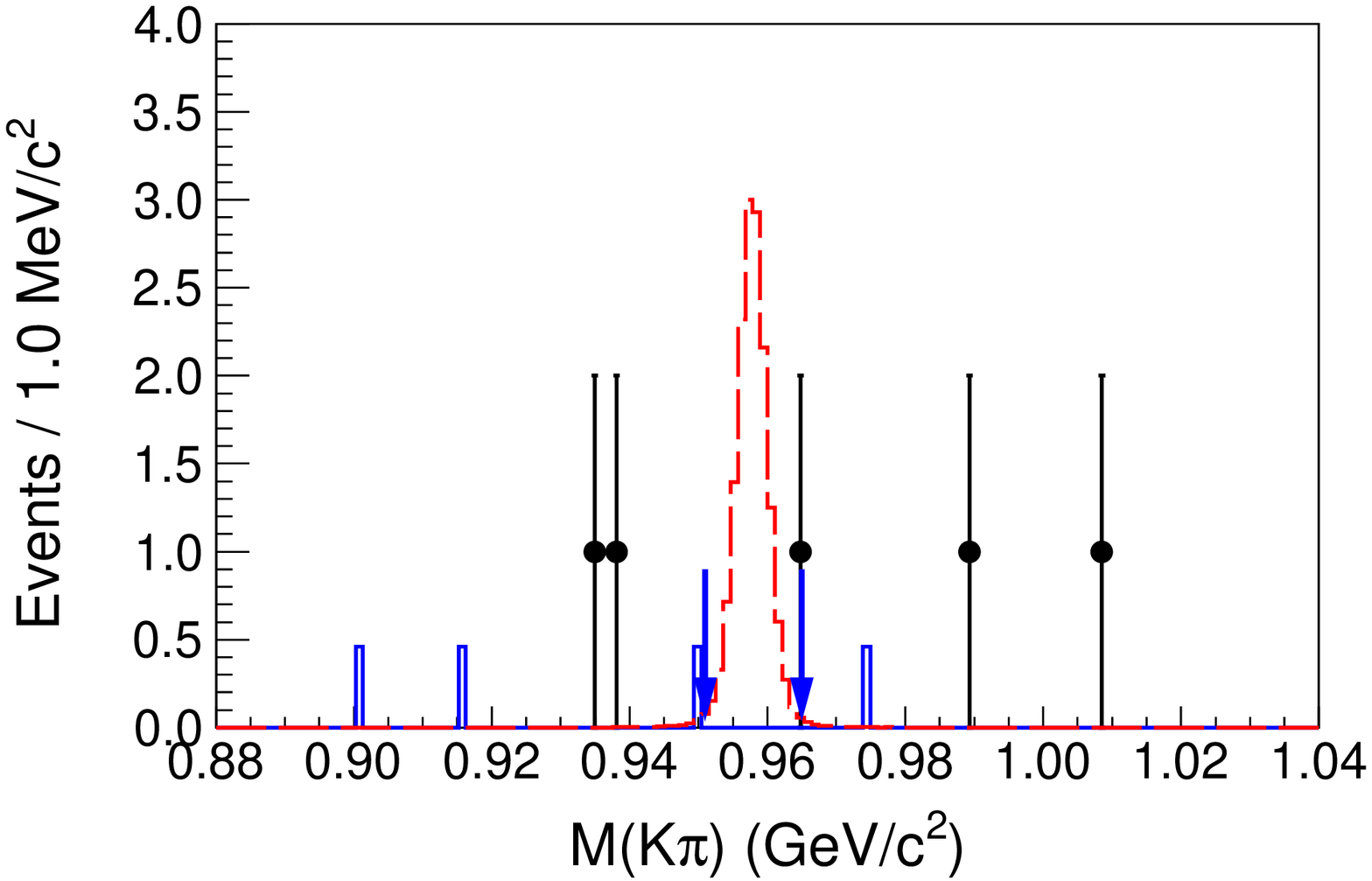}\put(-50,110){\bf (b)}
  \caption{\label{PhiKPi_plot}(a) Scatter plot of $M(\kk)$ versus
    $M(K^{\pm}\pi^{\mp})$, where the box indicates the signal region
    with $|M(\kk)-M(\phi)| < 15$~MeV/$c^2$ and
    $|M(K^\pm\pi^\mp)-M(\eta')| < 7$~MeV/$c^2$. (b) The $K^ \pm
    \pi^\mp$ invariant mass distribution, where the arrows show the
    signal region. The dots with error bars are for data, the dashed
    histogram is for the signal MC with arbitrary normalization, and
    the solid histogram is the background contamination from a MC
    simulation of $J/\psi \to \phi\pi^+\pi^-$.}
\end{figure*}
To search for $\eta'\to K^{\pm}\pi^{\mp}$, the two-body decay
$J/\psi\to\phi\eta'$ is chosen because of its simple event topology,
$K^+K^-K^{\pm}\pi^{\mp}$, and because the narrow $\phi$ meson is easy to
detect through $\phi\to K^+K^-$ decay.  The selection criteria for the
charged tracks are the same as that for the
$\jpsi\to\phi\eta', \eta'\to \gamma\pip\pim$ decay. Three tracks
are required to be identified as kaons with the combination of TOF and $dE/dx$
information and the remaining one is required to be identified as a pion.

A 4C kinematic fit imposing energy-momentum conservation is performed
under the $\kk K^\pm\pi^\mp$ hypothesis, and a requirement of
$\chi^2_{4C}<50$ is imposed.  To suppress the dominant background
contamination from $\jpsi\to\phi\pi^{+}\pi^{-}$, the $\chi^2_{4C}$ of the
$\kk K^{\pm}\pi^{\mp}$ hypothesis is required to be less than that for the
$\kk\pip\pim$ hypothesis. Candidates for $\phi\to K^+K^-$  are reconstructed from the
$K^{+}K^{-}$ combination with invariant mass closest to the nominal
mass value. The remaining kaon together with the pion form the $\eta'$
candidate.

Fig.~\ref{PhiKPi_plot} (a) shows the scatter plot to the invariant
mass $M(\kk)$ versus $M(K^\pm\pi^{\mp})$. The process $\phi\etap (\eta'\to K^{\pm}\pi^{\mp})$ would result in an enhancement of events around the nominal masses of
the $\phi$ and $\eta'$ mesons, while no evident cluster is
seen. Within three standard deviations of the $\phi$ mass,
$|M(\kk)-M(\phi)| < 15$ MeV/$c^2$, the $K^{\pm}\pi^{\mp}$ invariant
mass distribution is displayed in Fig.~\ref{PhiKPi_plot} (b); a
few events are retained around the $\eta^\prime$ mass region, shown as
the dots with error bars. To estimate the number of signal events
passing the selection criteria, a region of $\pm3\sigma$ around the
$\etap$ nominal mass is selected, that is
$|M(K^\pm\pi^\mp)-M(\eta')| < 7$~MeV/$c^2$, where $\sigma =2.2$~MeV
is the mass resolution determined from MC simulation. Only one event
survives in the signal region for further analysis.

To investigate the potential background contributions, a study with an inclusive MC
sample of $1.2\times 10^9$ generic $\jpsi$ decays is performed. It is found
that the remaining background events mainly come from
$J/\psi\to\phi\pi^+\pi^-$.  Therefore an exclusive MC sample of
$1.3\times 10^6$ $J/\psi\to\phi\pi^+\pi^-$ events is generated in
accordance with the partial wave analysis results of
Ref.~\cite{PhiPiPifromBESII}.
This sample corresponds to twice the
expected $\jpsi\to\phi\pip\pim$ events in data.  After normalizing to the
world average value for $\BR(\jpsi\to\phi\pip\pim)$, 2.0 events are expected
in the $K\pi$ mass range of [0.88, 1.04] GeV/$c^2$, with a total of 0.5
events in $\eta'$ signal region, as shown by the solid histogram in Fig.~\ref{PhiKPi_plot} (b).

To conservatively estimate the upper limit, it is assumed that the
only event in the signal region is a signal event. According to the Feldman-Cousins
method~\cite{UpperLimit}, the corresponding upper limit of the number
of events is $N^\text{UL} = 4.36$ at the 90\% C.L.

\section{Systematic Uncertainties}

The systematic uncertainties in branching fraction measurement
originate mainly from the differences of data and MC on tracking efficiency, photon reconstruction, PID
efficiency, and the 4C kinematic fit, different fitting
range and background shape,
uncertainties from $\BR(\phi\to\kk$) and $\BR(\eta'\to\gamma\pi^{+}\pi^{-})$, total number
of $\jpsi$ events and MC statistics. Other uncertainties related to the
common selection criteria of the channels
$\jpsi\to\phi\etap, \eta'\to K^{\pm}\pi^{\mp}$ and
$\jpsi\to\phi\etap, \etap\to\gamma\pip\pim$ cancel to first order in the
ratio between the branching fractions.

The systematic uncertainties associated with the tracking efficiency
and PID efficiency have been studied in the analysis of
$\jpsi\to p\bar{p}\pip\pim$ and
$\jpsi\to K_{S}^{0}K^{\pm}\pi^{\mp}$~\cite{trackingeff, PIDeff}. The
results indicate that the kaon/pion tracking and PID efficiencies for
data agree with those of MC simulation within 1\%.

The photon detection is estimated by the study of
$\jpsi\to\rho\pi$~\cite{trackingeff}. The difference in the detection
efficiency between data and MC is less than 1\% per photon, which is
taken as the systematic uncertainty because of the only photon in the
$\jpsi\to\phi\etap, \etap\to\gamma\pip\pim$ channel.

The uncertainty associated with the 4C kinematic fit comes from the
difference between data and MC simulation. The method used in this
analysis is to correct the tracking parameters of the helix fit to
reduce the difference between MC and data, as described in
Ref.~\cite{refsmear}. This procedure yields a systematic uncertainty
of 0.3\% and 1.0\% for the measurement of $\BR(\jpsi\to\phi\eta')$ and
the search of $\eta'\to K^{\pm}\pi^{\mp}$, respectively.

To estimate the systematic contribution due to the fit ranges,
several alternative fits in different ranges are performed. The
maximum difference on the number of signal events from alternative
fits in different mass ranges is 0.1\%, and this value is taken as
systematic uncertainty.  To estimate the systematic contribution due
to the background shape, a fit is performed replacing the 2nd-order
Chebychev polynomial function with an Argus function~\cite{ref:Argus};
the change of signal yields is found to be 0.04\%, which is negligible.

The decay $\jpsi\to\phi\eta', \phi\to\kp\km, \eta'\to\gamma\pip\pim$
is used as control sample to estimate the uncertainty from the $\phi$ mass
window criterion in the search of $\eta'\to K^{\pm}\pi^{\mp}$. The
$\phi$ mass window criterion is applied to the control sample, and a
fit is performed to $M(\gamma\pip\pim)$. After considering the efficiency difference,
the difference of 1.2\% in the number of
signal events between this fit and the nominal 2D fit is taken as the
uncertainty from the $\phi$ mass window.

The uncertainties on the intermediate-decay branching fractions of
$\phi\to\kk$ and $\eta'\to\gamma\pi^{+}\pi^{-}$ are taken from world
average values~\cite{PDG2014}.

The above systematic uncertainties together with the uncertainties due
to the number of $J/\psi$ events~\cite{JpsiNumberof2009, JpsiNumberofLiHuijuan} and MC
statistics are all summarized in Table~\ref{summary_of_syserr}, where
the uncertainties associated with MDC tracking, PID, branching
fraction of $\phi\to\kk$ cancel in the ratio
$\frac{\mathcal{B}(\eta^\prime \to K^\pm \pi^\mp) }
{\mathcal{B}(\eta^\prime \to\gamma\pi^+\pi^-) }$. The total systematic
uncertainty is taken to be the sum in quadrature of the individual
contributions.
\begin {table*}[htp]
  {\caption {Summary of systematic uncertainty sources and their contributions (in \%).}
    \label{summary_of_syserr}}
  \begin{tabular}{c|c|c}  \hline \hline
    Source & $\BR(\jpsi\to\phi\eta')$ & $\BR(\eta'\to K^{\pm}\pi^{\mp})$/$\BR(\eta'\to\gamma\pi^{+}\pi^{-})$\\ \hline
    Tracking efficiency & 4.0 & - \\ \hline
    PID efficiency & 4.0 & - \\ \hline
    Photon reconstruction & 1.0 & 1.0 \\ \hline
    4C kinematic fit & 0.3 & 1.0 \\ \hline
    Fit range & 0.1 & 0.1 \\ \hline
    Background shape & - & - \\ \hline
    $\phi$ mass window & - & 1.2 \\ \hline
    $\BR(\phi\to\kk)$ & 1.0 & - \\ \hline
    $\BR(\eta'\to\gamma\pi^{+}\pi^{-})$ & 2.0 & - \\ \hline
    $N_{\jpsi}$ & 0.8 & - \\ \hline
    MC statistic of $\eta'\to\gamma\pi^{+}\pi^{-}$ & 0.1 & 0.1 \\ \hline
    MC statistic of $\eta'\to K^{\pm}\pi^{\mp}$ & - & 0.1 \\ \hline
    Total & 6.2 & 1.9 \\ \hline \hline
  \end{tabular}
\end{table*}

\section{Results}
At the 90\% C.L., the upper limit on the ratio of
$\BR(\eta'\to K^{\pm}\pi^{\mp})$ to
$\BR(\eta'\to\gamma\pi^{+}\pi^{-})$ is given by
\begin{equation}
  \begin{aligned}
    \frac{\mathcal{B}(\eta'\to
      K^{\pm}\pi^{\mp})}{\mathcal{B}(\eta'\to\gamma\pi^{+}\pi^{-})}<\frac{
      N^\text{UL}\cdot\varepsilon_{\gamma\pi^+\pi^-} } {
      N_\text{sig}\cdot\varepsilon_{ K^\pm\pi^\mp}}\frac{1}
    {(1-\sigma_\text{syst})},
  \end{aligned}
\end{equation}
where $N^\text{UL}$ is the upper limit of the number of observed
events at the 90\% C.L. for $\eta^\prime\to K^\pm\pi^\mp$;
$\varepsilon_{ K^\pm\pi^\mp}$ and $\varepsilon_{\gamma\pi^+\pi^-}$ are
the detection efficiencies of $J/\psi\to\phi\etap$ for the two decays
which are obtained from the MC simulations; $\sigma_\text{syst}$ is the
total systematic uncertainty in the search of
$\eta'\to K^{\pm}\pi^{\mp}$.  The 90\% C.L. upper limit on the ratio
$\frac{\BR(\eta'\to
  K^{\pm}\pi^{\mp})}{\BR(\eta'\to\gamma\pi^{+}\pi^{-})}$ is determined
to be $1.3\times10^{-4}$ by using the values of different parameters
listed in Table~\ref{numbers_of_calculation}.
\begin {table}[htp]
  {\caption {Values used in the calculations of the branching
      ratios, including the fitted signal yields,  $N$ (or 90\%
      C.L. upper limit) and the detection efficiency, $\varepsilon$.}
    \label{numbers_of_calculation}}
  \begin {tabular}
    {cc c} \hline \hline
    Decay mode & $\varepsilon$ (\%) & $N$ \\
    \hline
    $\eta'\to K^{\pm}\pi^{\mp}$ & 36.75$\pm$0.04 & $<$4.36 (90\% C.L.)\\
    $\eta'\to\gamma\pip\pim$ & 32.96$\pm$0.04 & 31321$\pm$201 \\
    \hline \hline
  \end{tabular}
\end{table}

The branching fraction of $\jpsi\to\phi\eta'$ decay is calculated with the equation
\begin{equation}
  \begin{aligned}
    &\BR(\jpsi\to\phi\eta')\\
    &=\frac{N_\text{sig} /\varepsilon_{\gamma \pi^+\pi^-}}{ N_{\jpsi}\BR(\eta'\to\gamma\pi^{+}\pi^{-})\BR(\phi\to\kk)},
  \end{aligned}
\end{equation}
where $N_{\jpsi} = 1310.6\times10^{6}$ is the number of $J/\psi$ events
as determined by $J/\psi$ inclusive hadronic
decays~\cite{JpsiNumberof2009, JpsiNumberofLiHuijuan}.  The obtained value for the
branching fraction of $\jpsi\to\phi\eta'$ is
$(5.10\pm0.03(\text{stat.})\pm0.32(\text{syst.}))\times 10^{-4}$.

\section{Summary}
Based on the $1.3 \times 10^{9}$ $\jpsi$ events accumulated with the
BESIII detector, a search for the non-leptonic weak decay
$\eta'\to K^{\pm}\pi^{\mp}$ is performed for the first time through
the $\jpsi\to\phi\eta'$ decay. No evidence for
$\etap\to K^{\pm}\pi^{\mp}$ is seen, and the 90\% C.L. upper limit on
the ratio of
$\frac{\BR(\eta'\to
  K^{\pm}\pi^{\mp})}{\BR(\eta'\to\gamma\pi^{+}\pi^{-})}$ is measured
to be $1.3\times10^{-4}$. Using the world average value of
$\mathcal{B}(\etap\to\gamma\pi^+\pi^-)$~\cite{PDG2014}, the
corresponding upper limit on
$\mathcal{B}(\eta^\prime\to K^\pm\pi^\mp)$ is calculated to be
$3.8\times10^{-5}$.

For the determination of the ratio of
$\frac{\BR(\eta'\to
  K^{\pm}\pi^{\mp})}{\BR(\eta'\to\gamma\pi^{+}\pi^{-})}$, the
$\jpsi\to\phi\eta'$ decay with
$\phi\to\kk, \eta'\to\gamma\pi^{+}\pi^{-}$ is analyzed and the
corresponding branching fraction is
$\BR(J/\psi\to\phi\eta')=(5.10\pm0.03(\text{stat.})\pm0.32(\text{syst.}))\times10^{-4}$. It
is the most precise measurement to date and in agreement with the world average value.

\begin{acknowledgments}
The BESIII collaboration thanks the staff of BEPCII and the IHEP computing center for their strong support. This work is supported in part by National Key Basic Research Program of China under Contract No. 2015CB856700; National Natural Science Foundation of China (NSFC) under Contracts Nos. 11125525, 11235011, 11322544, 11335008, 11425524, 11105101, 11205117, 11575133, 11175189; the Chinese Academy of Sciences (CAS) Large-Scale Scientific Facility Program; the CAS Center for Excellence in Particle Physics (CCEPP); the Collaborative Innovation Center for Particles and Interactions (CICPI); Joint Large-Scale Scientific Facility Funds of the NSFC and CAS under Contracts Nos. 11179007, U1232201, U1332201, U1232109; CAS under Contracts Nos. KJCX2-YW-N29, KJCX2-YW-N45; 100 Talents Program of CAS; National 1000 Talents Program of China; INPAC and Shanghai Key Laboratory for Particle Physics and Cosmology; German Research Foundation DFG under Contract No. Collaborative Research Center CRC-1044; Istituto Nazionale di Fisica Nucleare, Italy; Ministry of Development of Turkey under Contract No. DPT2006K-120470; Russian Foundation for Basic Research under Contract No. 14-07-91152; The Swedish Research Council; U. S. Department of Energy under Contracts Nos. DE-FG02-04ER41291, DE-FG02-05ER41374, DE-FG02-94ER40823, DESC0010118; U.S. National Science Foundation; University of Groningen (RuG) and the Helmholtzzentrum fuer Schwerionenforschung GmbH (GSI), Darmstadt; WCU Program of National Research Foundation of Korea under Contract No. R32-2008-000-10155-0.
\end{acknowledgments}


\end{document}